\def\re#1{(\ref{#1})}
\def\beq{\begin{equation}}
\def\eeq{\end{equation}}
\def\beeq{\begin{eqnarray}}
\def\beeqn{\begin{eqnarray*}}
\def\eeeq{\end{eqnarray}}
\def\eeeqn{\end{eqnarray*}}

\def\m{\mu}
\def\n{\nu}

\newcommand{\DD}{{\cal D}}

\newcommand{\OO}{{\cal O}}

\newcommand{\WW}{{\cal W}}

\newcommand{\lp}{\left(}
\newcommand{\rp}{\right)}
\renewcommand{\lq}{\left[}
\renewcommand{\rq}{\right]}

\newcommand{\no}{\nonumber}
\newcommand{\ph}{\phantom}

\def\frac#1#2{ {{#1} \over {#2} }}

\def\ie{\hbox{\it i.e.}{ }}

\documentclass{JHEP3} 

\usepackage{epsfig,multicol}
\newcommand\fverb{\setbox\pippobox=\hbox\bgroup\verb}
\newcommand\fverbdo{\egroup\medskip\noindent%
                        \fbox{\unhbox\pippobox}\ }
\newcommand\fverbit{\egroup\item[\fbox{\unhbox\pippobox}]}
\newbox\pippobox

\title{On the invariance under area preserving diffeomorphisms of
noncommutative Yang-Mills theory in two dimensions}
\author{A. Bassetto, G. De Pol \\
Dipartimento di Fisica ``G.Galilei", Via Marzolo 8, 35131
Padova, Italy\\
INFN, Sezione di Padova, Italy\\
E-mail: \email{bassetto@pd.infn.it},~\email{depol@pd.infn.it}}
\author{A. Torrielli\\
Institut f\"ur Physik, Humboldt-Universit\"at zu
Berlin, \\Newtonstr. 15, D-12489 Berlin, Germany\\
E-mail: \email{torriell@physik.hu-berlin.de}}
\author{F. Vian\thanks{Supported by INFN, Italy, and partially by the European 
Community's Human Potential Programme under contract MRTN-CT-2004-005104~
``Constituents, fundamental forces and symmetries of the universe''}~\\
NORDITA, Blegdamsvej 17, DK-2100 Copenhagen \O, Denmark \\ 
E-mail: \email{vian@nbi.dk}}
\received{}             
\revised{}
\accepted{}             

\preprint{DFPD 05/TH/12 \\ HU-EP-05/12 \\ NORDITA-2005-22}        

\abstract{We present an investigation on the invariance properties
of noncommutative Yang-Mills theory in two dimensions under area preserving 
diffeomorphisms. Stimulated by recent remarks by Ambjorn, Dubin and Makeenko
who found a breaking of such an invariance, 
we confirm both on a fairly general ground and by means of 
perturbative 
analytical and numerical calculations that indeed invariance under area preserving 
diffeomorphisms is lost. However a
remnant survives, namely invariance under linear unimodular tranformations.}

        
\dedicated{}

\begin{document} 


\section{Introduction}

Ordinary Yang-Mills theory on a two-dimensional manifold (YM$_2$) has
been long known to be  
invariant under area preserving diffeomorphisms \cite{witten}. Such a 
symmetry of the classical action is apparent when one realizes that, 
in two dimensions, the action
depends only on the choice of the measure since the field strength of the 
Yang-Mills field is a two-form. Invariance then follows under 
all diffeomorphisms which preserve the volume element of the manifold.
The theory acquires an almost topological flavor \cite{m} and, as a 
consequence,
quantum YM$_2$ turns out to be exactly solvable. Beautiful pieces of 
literature have produced exact expressions for the partition function and 
Wilson loop averages exploiting such an invariance which led to the
discovery of powerful group-theoretic methods \cite{group}. 
In particular, the derivation of a regular version of the
Makeenko-Migdal loop equation by Kazakov and Kostov \cite{kazakov}
was possible in two dimensions due to the circumstance that the Wilson
loop average for a multiply-intersecting contour depends only on the
area of the windows it singles out on the manifold. This property can
also be explicitly verified via a perturbative computation in the
axial gauge with the principal value prescription, or on the lattice
as well, and finds its root in the above mentioned symmetry.

A perturbative computation with all-order resummation for the
Wilson loop average was performed  in \cite{sk}.
Invariance under area preserving diffeomorphisms allowed to choose a
particular contour, namely a circle, which provides a dramatic
simplification in calculations when the Wu-Mandelstam-Leibbrandt \cite{wmlbdg}
prescription for the light-cone gauge propagator is adopted.
The authors further probed the  invariance  performing  numerical
checks on contours with different shapes. Moreover, a
clarification of the role of 
non-perturbative contributions, and a generalization to contours with arbitrary 
winding numbers were then obtained in \cite{bassetto}, where again the 
forementioned
invariance turns out to play a crucial role.

So far the common lore has been that such an invariance persists in
Yang-Mills theories defined on a noncommutative two-dimensional space
and it should play a fundamental role inside the large noncommutative
gauge group underlying the model. The structure of such a group has been 
indeed widely investigated
because of the intriguing merging of internal and space-time
transformations. Its topology, from the very beginning, was argued to
be related to $U(\infty)$ and investigated, for example, in
\cite{ha}. The connection between the large-$N$ limit of $U(N)$
algebras and the algebra of area preserving reparametrizations of
two-dimensional tori was elucidated in \cite{ffz}.


A detailed study of the algebra of noncommutative gauge
transformations is contained in \cite{lsz}, where they are shown to generate
symplectic diffeomorphisms (see also \cite{dn}).

The complete gauge transformation is then shown to provide a deformation of
the symplectomorphism algebra of $R^d$.  
Finally, since symplectomorphisms in turn coincide  with  area preserving
diffeomorphisms in two dimensions, one might be led to expect that 
in this case noncommutative Yang-Mills could be exactly solved, exploiting 
the fact that observables depend purely on the area to apply powerful
geometric procedures. 

Most of the approaches to this problem start by contemplating the
theory on a noncommutative torus. An analysis of the gauge algebra on
such a manifold in connection with area preserving diffeomorphisms is
presented for example in \cite{s}. There one can exploit Morita equivalence
in order to relate the model to its dual on a commutative torus, in which case
 invariance under area preserving diffeomorphisms is granted.
A peculiar limit is 
then
needed to 
obtain the theory on the noncommutative plane, and one would still
expect that the same invariance is  there (see for instance \cite{ps2}).


Wilson loop computations directly on the noncommutative plane and in 
perturbation theory were performed in \cite{bnt1,bnt2,t}.
All the results obtained there were consistent with the expected invariance, 
namely the  expansion in the coupling
constant and  in $1/\theta$, $\theta$  being the
noncommutativity parameter, at the orders checked, was found to
depend solely on the area.

Recently, however, the authors of \cite{adm} were able to extend
the results of \cite{bnt1,bnt2} and found different
answers for the Wilson loop on a
circle and on a rectangle of the same area. Still, invariance
under rotations and symplectic dilatations was expected.

This is the main motivation to dwell on this issue and consider
the Wilson loop based on a wide class of contours with 
the same area. In this paper we limit ourselves to non self-intersecting contours
and find that, indeed,
invariance under area preserving diffeomorphisms is lost even for
smooth contours, the breaking being rooted in the non-local nature of the Moyal product.

A remnant of the invariance is nonetheless retained, precisely the invariance
under linear unimodular transformations ($SL(2,R)$).

The issue of volume preserving diffeomorphisms is also 
crucial in the study of dynamics of noncommutative D-branes
\cite{branes}. Since noncommutative Yang-Mills theory effectively describes 
 the low-energy behaviour of D-branes embedded in a constant 
antisymmetric background, the present analysis could be relevant also in a 
more general string-theoretic context.

The outline of the paper is as follows: In Sect.~2 
we prove that 
$SL(2,R)$ is preserved at any generic
order in axial-gauge perturbation theory, provided the gauge-fixing vector
is also consistently transformed. We  show how the procedure 
works at ${\cal{O}} (g^4)$, and how it can be easily generalized. 
In Sect.~3 
we show that the  $\theta^{- 2}$ term at ${\cal{O}} (g^4)$ is
invariant under $SL(2,R)$,
{\it without transforming the gauge vector}, by explicitly acting on it with the 
corresponding algebra generators (technical details are deferred to the Appendix). 
This term is used in Sect.~4 
to perform a 
series of 
analytic and numerical computations for a wide class of contours, both smooth 
and non-smooth, which confirm the outlined picture. In Section 5 we 
explain why the invariance under area preserving diffeomorphisms, which
occurs in ordinary YM$_2$, is broken to linear $SL(2,R)$ transformations in
the noncommutative case. 
Sect.~6 
is finally
devoted to our conclusions.
      
\section{Linear transformations: combined shape- and gauge-invariance in perturbation 
theory}

We will concentrate on the noncommutative Euclidean $U(1)$ gauge theory defined
on the two-dimensio\-nal plane. The results can be easily generalized to
the $U(N)$ case.
The quantum average of a Wilson loop can be defined by means of the
Moyal product as \cite{loop}~\footnote{We have omitted the base point 
of the loop and the related integration since the quantum average
restores translational invariance \cite{abou}.}  
\beq
\label{wloop}
\WW[C]=\int \DD (\tilde{n} A) \, e^{-S[\tilde{n} A]}  P_{\star} \exp \lp
ig \int_C (\tilde{n} A)
(z(s))\, \, \frac{(\tilde{n} \dot{z})}{\tilde{n}^2}ds \rp \,,
\eeq
where $C$ is a closed contour 
parameterized by $z(s)$, with $0 \leq s \leq 1$, $z(0)=z(1)$ and $P_\star$
denotes noncommutative path ordering along $z(s)$ 
from left to right
with respect to increasing $s$ of $\star$-products of functions, defined as
\beq
\label{Moy}
a * b = \left[\exp{[i \frac{\theta}{2} \epsilon^{\mu \nu}
{\partial}^{x_1}_{\mu} {\partial}^{x_2}_{\nu}]} a(x_1) b(x_2)\right]_{|x_1=x_2}.
\eeq
The axial
gauge fixing is $nA=0$, $n$ being an arbitrary, fixed vector, and the
vector $\tilde{n}$ can be chosen to obey
$\tilde{n}_{\mu}=\epsilon_{\mu \nu} n_{\nu}$,
so that $n^2=\tilde{n}^2$ and  $n \tilde{n} =0$.  
 
The perturbative expansion of $\WW [C]$,
expressed by Eq.~\re{wloop}, reads 
\beeq \label{loopert}
\WW   [C]&=&
\sum_{k=0}^\infty (ig)^k \int_{0}^1 ds_1 \ldots
\int_{s_{k-1}}^{1} ds_{k} 
\, 
\frac{\tilde{n} \dot{z}(s_1)}{\tilde{n}^2}
\ldots
\frac{\tilde{n} \dot{z}(s_k)}{\tilde{n}^2}
\no\\
&&\ph{
\sum_{k=0}^\infty (ig)^k } 
\times \langle 0\left|{\cal T}\lq  (\tilde{n} A)(z(s_1))
\star\ldots\star    (\tilde{n} A)(z(s_{k}))\rq \right|0\rangle, 
\eeeq
and is shown to be an (even) power series in $g$, so that we can write
\beq
\label{wpert}
\WW   [C]= 
\sum_{k=0}^\infty g^{2k}\WW_{2k} \, .
\eeq
The Moyal phase can be
handled in an easier way if we perform a Fourier transform, namely if
we work in momentum space. 
We use the axial gauge propagator $D_{\m\n}(p)$, which appears always
contracted with $\tilde{n}_\m \tilde{n}_\n$, 
so that the relevant
quantity is $\tilde{n}_\m \, D_{\m\n}(p)\,  \tilde{n}_\n=
n^2\tilde{n}^2/(np)^2$. Each order in the expansion Eq.~\re{wpert} 
is indeed
a sum of terms like the following 
\beeq
\label{order-2k}
&&\WW_{2k}^{\alpha \beta} = \int [ds] \, 
\tilde{n} \dot{z_1}
\ldots
\tilde{n} \dot{z_{2k}}
\int \frac{d^2p_1\ldots d^2p_k}{(2\pi)^{2k}}\,
\frac{1}{(np_1)^2\ldots(np_k)^2}\\
&&\ph{\WW_{2k} = \int [ds] \, 
\tilde{n} \dot{z}(s_1)
\ldots
\tilde{n} \dot{z}(s_{2k})}
\times \exp\{ i\sum_{i<j}^k M_{ij}\, p_i\wedge p_j\} 
\, \exp\{ i\sum_{r=1}^{k}
p_r\cdot (z_{\alpha(r)}-z_{\beta(r)})\}\,, \no
\eeeq
where $z_i$ is a shorthand notation for $z(s_i)$, $[ds]$ means the suitably 
ordered measure around the contour and we have taken into account that 
$n^4\,\tilde{n}^{-4}=1$. The notation
$\{\alpha(r)<\beta(r)\}$ refers to couples which exhaust the $2k$ positive integers and
$M_{ij}$ is
a suitable constant antisymmetric matrix. All these quantities depend on the topology
of the diagram considered and $p\wedge q \equiv \theta p_\m \epsilon_{\m\n} q_\n$.

\smallskip

One can now fairly easily show that the expression above is invariant 
under linear $SL(2,R)$ transformations provided we rotate the gauge vector accordingly, 
namely we
are considering combined gauge and linear unimodular deformation of the contour. In a
noncommutative setting these transformations belong to the $U(\infty)$ gauge 
invariance group.
Since we believe there is independence of the choice of the axial gauge-fixing vector, 
our result is tantamount to prove invariance under linear deformations of the contour. 

In the sequel we shall explicitly exhibit the invariance of the fourth order 
term ${\cal O}(g^4)$, 
but the proof can be straightforwardly
generalized to any arbitrary order, 
{\it i.e.} to the generic term appearing in Eq.~(\ref{order-2k}).
Eq.~\re{order-2k} leads to the following expression for ${\cal W}_4^{np}$, the non-planar
contribution to the Wilson loop at ${\cal O}(g^4)$
\beq
\label{order-4}
\WW_4^{np} = \int [ds] \, 
\tilde{n} \dot{z}_1
\ldots
\tilde{n} \dot{z}_4
\int \frac{d^2p\,d^2q}{(2\pi)^{4}}\,
\frac{\exp i \{ p\wedge q + p(z_1-z_3)+ q(z_2-z_4)\}}{(np)^2\,(nq)^2}
\,. 
\eeq

We now perform on  $z_\mu$ the transformation
$z_\mu = S_{\mu\nu} \, \xi_\nu$,
with $\det S = 1$, corresponding to the linear unimodular deformation
of the contour discussed above.
We simultaneously introduce the vectors $P_\m$, $Q_\n$ such that  $p_\mu= (P
S^{-1})_{\m}$, $q_\n= (Q S^{-1})_{\n}$. It is immediate to verify that
$p\wedge q=P\wedge Q$,
since the following matrix relation holds 
\beq
\label{s-matrix}
S^{-1}\epsilon (S^{-1})^T=\epsilon \,,
\eeq
where the matrix $\epsilon$ is defined as $(\epsilon)_{\m\n}=\epsilon_{\m\n}$.
Then Eq.~\re{order-4} becomes
\beq
\label{order-4trans}
\WW_{4}^{np} = \int [ds] \, 
(\tilde{n}S \dot{\xi})_1
\ldots
(\tilde{n} S\dot{\xi})_4
\int \frac{d^2P\,d^2Q}{(2\pi)^{4}}\,
\frac{\exp i \{ P\wedge Q + P(\xi_1-\xi_3)+
Q(\xi_2-\xi_4)\}}{(PS^{-1}n)^2\,(QS^{-1}n)^2} \,. 
\eeq
The next step consists in defining a new gauge fixing vector according
to $\nu = S^{- 1} n$ and choosing $\tilde{\n}$ such that $\tilde{\n} 
= \tilde{n} S$. Hence we still have $\tilde{\nu}=\epsilon \nu$ and one
can easily realize, with the help of Eq.~\re{s-matrix}, 
that  the conditions $\nu \tilde{\nu}=0$ and $\tilde{\nu}^2 = \nu^2$
are satisfied. Finally, substituting $n$, $\tilde{n}$ with $\n$,
$\tilde{\n}$, respectively, in Eq.~\re{order-4trans}, we conclude our
proof, namely that the Wilson loop average, at $\OO
(g^4)$, is invariant under {\it linear}
$SL(2,R)$ transformations provided we rotate the gauge vector accordingly.

\section{$SL(2,R)$ invariance of the coefficient of $\theta^{-2}$ at ${\cal O}(g^4)$}

We prove the invariance under linear deformations of the contour 
{\it without changing the gauge vector} for the $\theta^{-2}$ term in the 
expansion 
in $1\over \theta$ of the Wilson loop at ${\cal O}(g^4)$.
This quantity will in turn be used to prove the breaking of the {\it local}
unimodular invariance in the noncommutative context.

The $\theta^{-2}$ term of the Wilson loop, at the perturbative order 
$g^4$, which will be indicated as 
$W[C]$ throughout this section for the sake of simplicity, 
will be computed below \rm
for a number of 
various contours, in order to check whether, and to what extent, invariance 
under area preserving diffeomorphisms holds. Invariance would imply  that 
$W[C]=k A_C^4$, 
where $A_C$ is the area enclosed by the contour $C$,
the constant $k$  being universal, {\it i.e.}
independent of the shape of $C$.
The starting point for our explicit (both analytical and numerical) 
computations is \cite{bnt1,adm}

\beeq \label{makesing}
&&W[C]=\frac{g^4}{4! 4 \pi^2 \theta^2}\times \\&&\times 
{\mbox{\bf P}}\!
\int\!\int\!\int\!\int 
\frac{((x_1 -z_1)(y_2-t_2)-(y_1-t_1)(x_2-z_2))^4}{(x_2-y_2)^2(y_2-t_2)^2}
dx_2 dy_2 dz_2 dt_2\,,\nonumber 
\eeeq
where the integration
variables are ordered as  $x<y<z<t$ with 
respect to a given parametrization of $C$ and the subscripts $1,2$ refer to 
the Euclidean components of the coordinates (the axial gauge is $A_1=0$).

This expression 
can be seen to be equivalent to \footnote{Yuri Makeenko, private communication.}
\beeq \label{makereg}
W[C]= && \frac{g^4}{4! 4 \pi^2 \theta^2}\times \left[ A_C^4 +\right. 
\nonumber \\
&& + 30\, {\mbox{\bf P}}\!\int \!\int \!\int 
x_1 y_1 z_1 (x_1(y_2-z_2)+y_1(z_2-x_2)+z_1(x_2-y_2))\,dx_2 \,dy_2 \,dz_2 +
\nonumber \\ 
&& +\frac{5}{2}
\oint \oint \left(\frac{4}{3}x_1^3 y_1 +x_1^2 y_1^2 +\frac{4}{3}x_1
y_1^3\right) (x_2-y_2)^2\,
dx_2 \,dy_2\,\left. \right]\, ,
\eeeq
which is particularly suitable for performing  calculations since 
denominators no longer appear. 
Here the triple integral is ordered according to
$x<y<z$, while the double integral is not ordered.
Another obvious advantage of Eq.~\re{makereg} is that, being explicitly 
finite, it can be run as it stands by a computer program, in order to 
estimate numerically $W[C]$ for curves that are analytically 
untractable, such as the even order Fermat curves $C_{2n}\equiv \{(x,y):\ x^{2n}+ 
y^{2n}\ =1$ {\rm with}  $n\ge 1\}$.  

The result, as we shall discuss in detail, is that invariance is lost, but a 
weaker remnant still holds, namely invariance under linear area 
preserving 
maps (elements of $SL(2,R)$). Invariance of the quantum average of a Wilson loop 
under translations 
is automatic in noncommutative theories owing to the trace-integration over space-time
(see also \cite{abou}).
If we define 
\beq\label{remnant}
W[C]=\frac{g^4 A_C^4}{4! 4 \pi^2 \theta^2} \, {\cal I}[C]\, ,
\eeq
it is 
apparent from Eq.~(\ref{makesing}) 
and from dimensional analysis
that $\cal I [C]$ is  dimensionless and  
characterizes the {\em shape} (and, {\it a priori}, the orientation) of a 
given contour. 

Before reporting on specific calculations it is worth showing explicitly that 
Eq.~(\ref{makesing})
is invariant under linear area preserving maps. This turns out to be more 
involved than expected. 

A convenient choice for the  infinitesimal generators of area preserving 
diffeomorphisms is given by the set of analytical vector fields 

\beq \label{apdgen}
\{ V_{m,n}\equiv n x_1^m x_2^{n-1} \partial_{x_1}- m x_1^{m-1}x_2^n
\partial_{x_2}\,; (m,n)\in \mathbb{N}\times \mathbb{N}-(0,0)\},
\eeq

which close on the infinite-dimensional Lie algebra
\[
\left[V_{m,n},V_{p,q}\right]= (np-mq) V_{m+p-1,n+q-1}\,.
\]
When acting on the coordinates, these generators produce the infinitesimal 
variations
\[\delta_{V_{m,n}} (x_1,x_2)=(n x_1^m x_2^{n-1},-m
x_1^{m-1}x_2^n)\,,\]
which obviously satisfy the (infinitesimal) unimodularity condition 
$\partial_i \delta x_i=0$.
We notice that the 
generators with $n+m \le 2$ span a finite subalgebra and
exponentiate to unimodular inhomogeneous linear maps,
namely  translations and elements of $SL(2,R)$,
while  higher level generators exponentiate to 
non-linear 
area preserving maps.

Eq.~(\ref{makesing}) 
is easily seen to be 
invariant under translations (which are built in) 
and under $V_{0,2}, V_{1,1}$: the numerator 
is invariant under linear symplectic maps, and, as far as denominator and measure are
concerned, $V_{0,2}$ does not affect the component parallel to the axial gauge 
vector, while $V_{1,1}$ scales it homogeneously. Indeed Eq.~(\ref{makesing}) 
is manifestly invariant under the subgroup generated by $V_{0,2}, V_{1,1}$, namely 
under transformations of the type

\[
x_1'=a x_1 +b x_2\,,\qquad x_2'=a^{-1} x_2\,.
\]

A proof of invariance under a third, 
independent generator, the simplest choice being  $V_{2,0}$, can be obtained 
via the equivalent form Eq.~(\ref{makereg}) 
and is presented in the Appendix.

\section{Explicit computations for various contours}

The computation of the Wilson loop is in principle straightforward for polygonal
contours, since only polynomial integrations are required; nonetheless,
a considerable  amount of algebra makes it rather
involved. 

Here we summarize our results:

\begin{itemize}

\item{{\bf Triangle:}} ${\cal I}[\mbox{Triangle}]$ was computed for an 
arbitrary triangle, the result being ${\cal I}[\mbox{Triangle}]=
\frac{8}{3}\simeq 2.6667$. This 
is  consistent with $SL(2,R)$ invariance, 
since any two given  triangles of equal area can be mapped into each
other  via a linear unimodular map.

\item{{\bf Parallelogram:}} ${\cal I}[\mbox{Parallelogram}]$ was computed for an 
arbitrary parallelogram
with a basis along $x_1$, the result being
${\cal I}[\mbox{Parallelogram}]=\frac{91}{36}\simeq 2.5278$. Again, this is 
consistent with $SL(2,R)$ invariance, by the same
token as above.

\item{{\bf Trapezoid:}} Here we can see analytically an instance of the
broken invariance.
Trapezoids of equal area cannot 
in general be mapped into one other by a linear 
transformation, since the ratio of the two basis $b_1/b_2$ is a $SL(2,R)$ 
invariant. 
One might say that the space of trapezoids of a given area, modulo  $SL(2,R)$,
has at least (and indeed, exactly) one modulus, which can  be conveniently 
chosen as the ratio $b_1/b_2$. Actually, the result we obtained for  
${\cal I}[\mbox{Trapezoid}]$ reads
\beq \label{trapezoid}
{\cal I}[\mbox{Trapezoid}]= \frac{4(6 b_1^4 +24 b_1^3 b_2 +31 b_1^2 b_2^2 +
24 b_1 b_2^3 +6 b_2^4)}{9(b_1+b_2)^4}\,,
\eeq
namely a function of $b_1/b_2$ only, duly invariant under the
exchange of $b_1$ and $b_2$, 
and correctly reproducing ${\cal I}[\mbox{Parallelogram}]$ when $b_1=b_2$, and 
${\cal I}[\mbox{Triangle}]$ when $b_1=0$. It is plotted in Fig.~\re{trap1}.

\FIGURE[h]{\epsfig{file=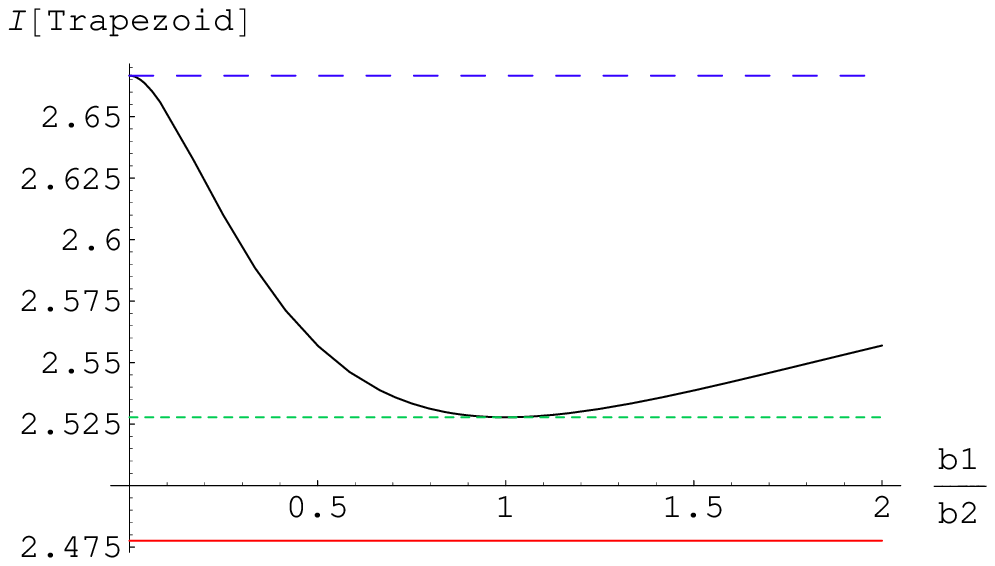}\label{trap1}\caption{
${\cal I}[\mbox{Trapezoid}]$ as a function of the ratio of the basis
$b_1/b_2$; the continuous, dashed and dotted straight lines refer to
${\cal I}[\mbox{Circle}]$, ${\cal I}[\mbox{Triangle}]$ and ${\cal I}[\mbox{Square}]$, respectively.}}

\end{itemize}

Thus, the main outcome of our computations is that different polygons
turn out to produce different results, unless they can be mapped into
each other through linear unimodular maps.

As opposed to polygonal contours, smooth 
contours  cannot be in general computed analytically, with the noteworthy 
exceptions of the circle and the ellipse. A circle can be mapped to any 
 ellipse of equal area by the forementioned  area preserving linear
maps. Indeed, it is easy to realize from Eq.~(\ref{makereg}) (because of 
homogeneity, using the obvious 
parametrizations), that an ellipse with the axes
parallel to $x_1,x_2$ gives the same result as the circle
\beq \label{circle}
{\cal I}
[\mbox{Circle}]={\cal I}[{\mbox{Ellipse}}]=1 +\frac{175}{12 \pi^2} \,
\eeq
 and it turns out to be the lowest value among the (so far computed)
non self-intersecting contours. It might be interesting to understand
whether this circumstance has any deeper origin.
Clearly the same equality does not hold for polygonal contours, {\it e.g.}
${\cal I}[{\mbox{Triangle}}]\neq {\cal I}[{\mbox{Circle}}]$. 

In the lack of explicit 
computations for smooth contours, different from circles and ellipses, this 
scenario might have left open the question whether in the
noncommutative case invariance would
still be there for smooth contours of equal area, 
only failing for polygons  due to the presence of cusps. We did the check by
numeric computations  for the (even order) 
Fermat curves  $C_{2n}$ which constitute a family of closed and smooth contours,
``interpolating'', in a discrete sense, between two analytically known 
results, \ie  the circle ($n=1$) and the square (the $n\to\infty$ limit). The 
numerical computations were performed with two different and independent 
algorithms, which provided identical results within the level of accuracy required. 

As shown in Fig.~(\ref{fermatcurves}) and in Table~(\ref{mytable}),
${\cal I}[C_{2n}]$  definitely varies with $n$
and in the 
$n\to\infty$ limit approaches ${\cal I} [{\mbox{Square}}]$. 

Thus 
we conclude that   invariance under area preserving
diffeomorphisms does 
{\em not} hold.
\FIGURE[h]{\epsfig{file=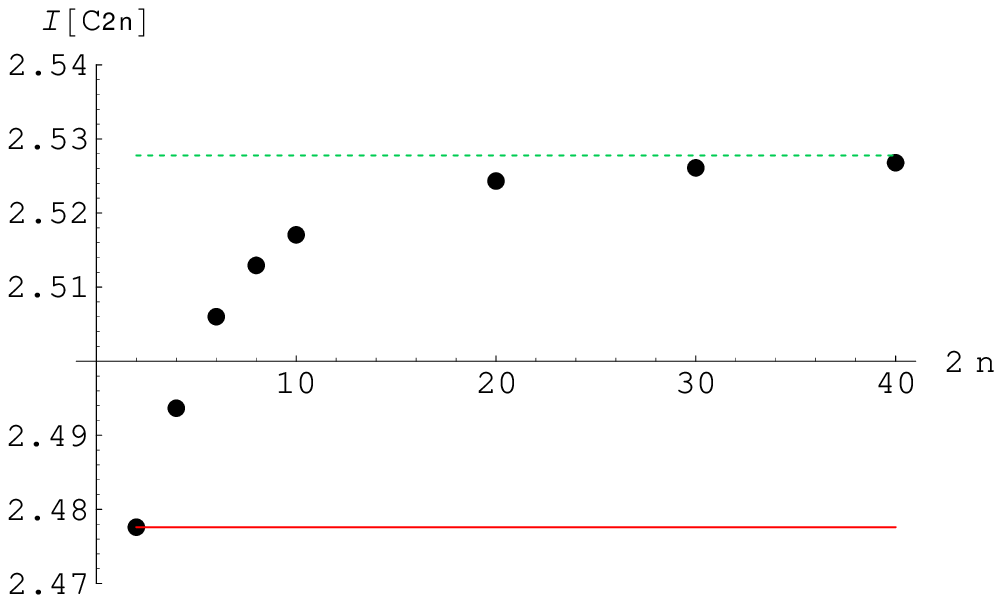}\label{fermatcurves}
\caption{${\cal I}[C_{2n}]$ for Fermat curves with different $n$; the
continuous, dotted lines refer to ${\cal I}[\mbox{Circle}]$, ${\cal
I}[\mbox{Square}]$, respectively.}}
\TABLE[h]{
\label{mytable}
\begin{tabular}[!h]{|c|c|c|c|c|c|c|c|c|}\hline 
 n & 1 & 2 & 3 & 4 & 5 & 10 & 15 & 20 \\ \hline
 ${\cal I}[C_{2n}]$ & 2.4776  & 2.4937 & 2.5060 & 2.5129 & 2.5171 
& 2.5243 & 2.5361 & 2.5268  \\ \hline
\end{tabular}
\caption{Numerical results for ${\cal I}[C_{2n}]$ for Fermat curves
with different $n$.} 
}

\vskip .5truecm

Furthermore, we explicitly checked that the numerical results displayed above
are invariant
(within our numerical accuracy) under rotations of the contours.

\smallskip

By considering higher order generators we were able to show that:
\begin{enumerate}
\item
All the generators annihilate $W[C]$ on a circle: 
$\delta_{V_{m,n}}W[C]|_{\mbox{Circle}}=0\,\, \forall(m,n)\,.$ 
This seems consistent with the fact that $W[\mbox{Circle}]$ is the lowest 
value among the so far checked non self-intersecting contours, and suggests 
that actually the 
circle is likely to be a {\em local} weak minimum (of course, there must be 
flat directions, corresponding to the surviving mentioned symmetries). 
Nevertheless, the constrained second variation is not easy to 
compute, and we have not succeeded so far to see whether it is positive 
semi-definite as we would expect. 
\item

It was checked by explicit computation that higher level 
generators do not leave $W[C]$ invariant around a generic triangle. 
As an example, the variation under $V_{3,0}$ 
of the triangle defined in the $(x_1,x_2)$ plane by the vertices 
$\{(0,0),(l_1,h_1),(l_2,h_2)\}$ is
\[
A_C^4 \cdot \delta_{V_{3,0}} {\cal I}[C]|_{\mbox{Triangle}}=
\frac{1}{12}l_1 l_2(l_1-l_2)(l_1 h_2 -l_2 h_1)^3
\]
and vanishes (for non degenerate triangles) only if one of the sides is 
parallel to the $x_2$ axis. Different 
vanishing conditions arise for different generators, and it can be seen that 
for no triangle at all is $W[C]$ invariant under the full group of area 
preserving diffeomorphisms.
\end{enumerate}

\section{Arguments concerning invariance under general transformations}

Let us provide now a general argument which shows that the expectation value 
of the Wilson loop is invariant under area preserving diffeomorphisms in 
ordinary YM$_2$. We discuss for simplicity the $U(1)$ case.

To start with, let us consider the  path integral in $d$ dimensions
\beq
\label{defi}
W[C]=\int {\cal D}A e^{-S[A]} w[C,A] \,,
\eeq
where $S[A]=\frac1{4}\int F_{ij}F_{kl}\eta^{ik}\eta^{jl} d^d x$ is a functional of 
the vector field $A$, and $w[C,A]=P \exp{i\, \int_C A_i dx^i}$ is a functional of 
$A$ and of the contour $C$. 
This has been formulated in cartesian coordinates with metric $\eta_{ij}$.
Under a different
choice of coordinates $x'=x'(x)$, $W[C]$ can be 
rewritten as  
\beq
\label{lop}
W[C]=\int {\cal D} A e^{-S_{gen}[A',\,g']} w[C',A'],
\eeq
where $S_{gen}[A,g]=\frac1{4}\int F_{\mu \nu} F_{\rho \sigma} g^{\mu \nu} 
g^{\rho \sigma} \sqrt{\det{g}} \,d^d x$, 
provided $A$ and $\eta$ transform to
$A',g'$ like tensors. Notice that $\det{g}$ is positive and the 
definition of $F_{\mu \nu}$ is left 
unchanged in the covariantized formulation.  

On the other hand, we can
consider the same functional computed for the deformed contour $C'$
\beq
\label{nlop}
W[C']=\int {\cal D}A e^{-S[A]} w[C',A],
\eeq
the deformation being described by the same map $x'=x'(x)$ as above.

The condition
\beq
\label{sym}
S_{gen}[A,g']=S[A]
\eeq
would describe a symmetry
of the classical action,
which, in dimensions $d>2$, is a tight condition to fulfill. In $d=2$, due to 
the circumstance that $F_{\mu \nu}$ is a two-form, we get 
\beq
\label{ac}
S[A]=\frac1{2} \int F_{12}^2 d^2x,
\eeq
while in $S_{gen}[A,g]$ the contractions with the inverse metric contribute a 
factor
$(\det g)^{-1}$
\beq
\label{gen}
S_{gen}[A,g]=\frac1{2} \int F_{12}^2 \frac{1}{\sqrt{\det{g}}}d^2x .
\eeq
The condition Eq.~(\ref{sym}) 
then amounts to $\det{g}=1$, which is ensured if the 
Jacobian of the map is one,
namely
if $C$ can be deformed to $C'$ by an area preserving map.

It is well known that in ordinary YM$_2$ a Wilson loop depends only
on the area it encloses
(and not on its shape); thereby the classical symmetry
persists
at the quantum level, in turn implying ${\cal D}A = {\cal D}A'$.

\smallskip

When we turn to the noncommutative theory in  $d=2$,
the expectation value of the 
Wilson loop becomes 
\beq
\label{ncw}
W_{nc}[C,*]=\int {\cal D}A e^{-S_{nc}[A,*]} w_{nc}[C,A,*]\,,
\eeq
where the Moyal product has been introduced
\beeq
&&S_{nc}[A,*]=\frac1{4}\int F_{ij}*F_{kl}\,
\eta^{ik}\eta^{jl} d^d x\no \\
&&w_{nc}[C,A,*]=P_{*} \exp{i\, \int_C A_i dx^i}\no \,,
\eeeq
and 
the dependence of the involved functionals on $*$ is explicitly exhibited in 
order to make what follows clear.

Also $W_{nc}$ can be rewritten 
in general coordinates, provided a 
covariantized $*^{g}$ product is defined as
\beq
\label{gm}
a *^g b = \left[\exp{[i \frac{\theta}{2} \frac{\epsilon^{\mu \nu}}{\sqrt{\det{g}}} 
{\cal D}^{x_1}_{\mu} {\cal D}^{x_2}_{\nu}]} a(x_1) b(x_2)\right]_{|x_1=x_2},
\eeq
${\cal D}_{\mu}$ being the covariant derivative associated to the Riemannian 
connection for the metric 
$g$.~\footnote {We stress that, by introducing $*^g$, we are 
not formulating the theory on a curved space. Instead,
we are just 
rewriting the theory on the flat space
in general coordinates. It should be evident, from 
general covariance of the tensorial quantities involved, that
Eq.~(\ref{gm}) 
in Cartesian 
coordinates reproduces the usual Moyal product Eq.~(\ref{Moy}). 
Notice also that,  
since by definition
${\cal D}_{\mu}g_{\rho \sigma}(x)=0$, it is irrelevant to choose either $x_1$
or $x_2$ as argument of $\det g(x)$ 
in Eq.~(\ref{gm}), and, by the same token, 
$*^g$ is uneffective 
when acting on the metric tensor  $g$. 
 It may also be worth 
noticing that the commutativity of covariant derivatives in flat space would allow to 
prove independently the associativity of $*^g$.}
Under the same reparametrization we then obtain
\beq
\label{rep}
W_{nc}[C,*]=\int {\cal D} A e^{-S_{nc, gen}[A',g',*^{g'}]} w_{nc}[C',A',*^{g'}],
\eeq
where the noncommutative action in general coordinates 
\beq
\label{ncS}
S_{nc, gen}[A,g,*^g]=\frac1{4}\int F_{\mu \nu} *^g F_{\rho \sigma} g^{\mu \nu} 
g^{\rho \sigma} \sqrt{\det{g}} \,d^2 x
\eeq
and Wilson loop 
\beq
\label{Wil}
w_{nc,gen}[C,A,*^g]=P_{*^g} \exp{i\, \int A_{\mu}dx^{\mu}}
\eeq
 have been introduced. 

Assuming the absence of functional anomalies also in the noncommutative case,
we compare Eq.~(\ref{rep}) 
with the 
functional $W_{nc}[C',*]$ computed for the deformed contour $C'$
\beq
\label{def}
W_{nc}[C',*]=\int {\cal D} A e^{-S_{nc}[A,*]} w_{nc}[C',A,*].
\eeq
The two quantities coincide if the following two sufficient conditions
are met
\begin{itemize}
\item $*^g=*$, 
\item
$S_{nc,gen}[A,g',*]=S_{nc}[A,*].$
\end{itemize}

These conditions imply that the map is at most linear, since the 
Riemannian
connection must vanish, and that 
its Jacobian 
equals unity. In conclusion, only $SL(2,R)$ linear maps are allowed.

\section{Conclusions}

Gauge theories defined on a noncommutative two-dimensional manifold  were from the
 beginning believed to be invariant under the group of area preserving 
diffeomorphisms, as it occurs in their commutative counterparts. 
This property holding, one would have a deeper 
understanding of the structure of the noncommutative gauge group underlying these 
models, and a larger set of geometric tools
available for challenging a complete solution. However, a perturbative computation 
of the Wilson loop revealed a lack of
this invariance: at order ${\cal{O}} (g^4)$ in the coupling constant 
and ${\cal{O}} (\theta^{- 2})$ in the  noncommutativity
parameter, the Wilson loop $W[C]$ is not the same when evaluated on a circle 
and on a rectangle of equal area \cite{adm}. While this could be the signal 
of a breaking 
of 
invariance under the group of area preserving diffeomorphisms 
at the 
perturbative level, the questions of the 
generalization of this result to other classes of contours, and of the possible 
existence of unbroken subgroups, immediately arose.
    
In this paper we  reported in detail a number of analytic and numerical computations 
of the above mentioned term $W[C]$
 for a wide choice of different contours. In so doing we  confirmed the 
 breaking of the invariance: we  found 
different values for triangles, parallelograms, trapezoids, circles and Fermat curves. 
An interesting  result we  found is that the 
latter nicely interpolate between the values of the 
the circle and of the square. 
We pointed out
the existence of an unbroken subgroup, namely the one of  
area preserving  linear maps of the plane $SL(2,R)$. 
We also explained 
why invariance under a local area preserving diffeomorphism,
 which is present in the commutative case, cannot 
persist in the noncommutative context, owing to the non-local nature of the Moyal
product, at least perturbatively. 

While our findings give a generalized proof of the perturbative breaking of 
  invariance under area preserving diffeomorphisms, some issues
deserve further investigation. It would be interesting to possibly extend our results
to nonperturbative approaches 
appeared in the literature \cite{nonpert}. In turn they might also be relevant
 for the physics of membranes, which is 
tightly related to noncommutative 
gauge theories. On another side they may entail
important consequences on the analysis of the merging of space-time and internal 
symmetries in a 
noncommutative context.

\section{Acknowledgements}
We thank Jan Ambjorn and Yuri Makeenko for  
stimulating insights stressing the breaking of the general invariance under 
area preserving diffeomorphisms in noncommutative YM$_2$.
We are grateful to Pieralberto Marchetti for general comments concerning 
invariance properties in QFT.
Finally, A.T. wishes to thank 
Harald Dorn, Christoph Sieg  and Sebastian de Haro for discussions.

\section{Appendix}

We show here that Eq.~(\ref{makereg}) in the main text is invariant under the action
of the infinitesimal generator $V_{2,0}$, which can be read from Eq.~(\ref{apdgen}). 
The proof goes as follows. First we compute the variation of
Eq.~(\ref{makereg}) 
under a generic infinitesimal tranformation  
$(\delta x_1,\delta x_2)$. Let us indicate the integrands in the triple and double
integrals in Eq.~(\ref{makereg}) as $f(x,y,z)$ and $g(x,y)$  and notice 
that $f$ is a completely antisymmetric function of its arguments, whereas 
$g$ is symmetric. Integrating 
by parts and taking into due account the fact that the curve is closed, which, 
together with the aforementioned symmetry properties, allows to cancel all 
the
``finite parts'', we find
\beeq
&&A_C^4 \cdot \delta {\cal I}[C] = 
 \frac{5}{2}\oint \oint[\partial_{x_1}g
(\delta x_1 \, dx_2 dy_2-\delta x_2 \, dx_1 dy_2)+ \partial_{y_1}g 
(\delta y_1 \, dx_2 dy_2-\delta y_2 \, dx_2 dy_1) ] \nonumber
\\ 
&&+ 
30 \, 
{\mbox{\bf P}} \!\int \!\!\!\int \!\!\!\int 
[\partial_{x_1}f
(\delta x_1 \, dx_2 dy_2 dz_2 - \delta x_2 \, dx_1 dy_2 dz_2) 
+\partial_{y_1}f 
(\delta y_1 \, dx_2 dy_2 dz_2 - 
\delta y_2 \, dx_2 dy_1 dz_2)\no \\ 
&&
\ph{30 \, 
{\mbox{\bf P}} \!\int \!\!\!\int \!\!\!\int}
+ \partial_{z_1}f 
(\delta z_1 \, dx_2 dy_2 dz_2 -\delta x_2 \, dx_2 dy_2 dz_1) ] \,;
\eeeq

when we specialize the expression above to the $V_{2,0}$ generator, it becomes

\beeq
\label{variation}
&& A_C^4 \cdot \delta_{V_{2,0}} {\cal I}[C] = 
5 \oint \oint [(4 x_1^3 y_1+ 2x_1^2 y_1^2+\frac{4}{3}x_1 y_1^3)
(x_2^2-2x_2 y_2+y_2^2)\,dx_1 dy_2 \\
&&
\ph{ A_C^4 \cdot \delta_{V_{2,0}} {\cal I}[C] =5 \oint [ }
+(\frac{4}{3} x_1^3 y_1+ 2x_1^2 y_1^2+4 x_1 y_1^3)
(x_2^2-2x_2 y_2+y_2^2)\,dx_2 dy_1] \nonumber  
\\ 
&&+60 \, {\mbox{\bf P}}\!\int\!\!\!\int\!\!\!\int [ 
(2x_1^2 y_1 z_1 (y_2-z_2)+ x_1 y_1^2 z_1 (z_2-x_2) +
x_1 y_1 z_1^2 (x_2 -y_2))\, dx_1 dy_2 dz_2 \nonumber \\
&&\ph{+60 \, {\mbox{\bf P}}\!\int\!\!\!\int\!\!\![}
+(x_1^2 y_1 z_1 (y_2-z_2)+ 2x_1 y_1^2 z_1 (z_2-x_2)+ 
x_1 y_1 z_1^2 (x_2 -y_2))\, dx_2 dy_1 dz_2\nonumber \\
&&
\ph{+60 \, {\mbox{\bf P}}\!\int\!\!\!\int\!\!\![}
+(x_1^2 y_1 z_1 (y_2-z_2)+ x_1 y_1^2 z_1 (z_2-x_2)+ 
2 x_1 y_1 z_1^2 (x_2 -y_2))\, dx_2 dy_2 dz_1] \,.\nonumber 
\eeeq

A few tricks now are useful to simplify Eq.~(\ref{variation}), namely:
\begin{enumerate}
\item
terms in which only one component of one of the points $x,y,z$ appears,
such as $x_1^2 dx_1$, can be explicitly integrated, giving rise to either 
ordered double integrals (from the triple integral) or simple  loop integrals 
(from the double one). That is, {\it e.g.}
\[
\mbox{\bf P}\!\int \!\int \!\int
y_1 y_2 z_1^2 x_1 dx_1 dy_2 dz_2 \to 
\frac{1}{2}\mbox{\bf P} \!\int \!\int 
y_1 y_2(y_1^2 -t_1^2)z_1^2 dy_2 
dz_2\,,
\]
$t=(t_1,t_2)$ being the starting point of the parametrization;

\item
terms in which each point appears with 
both components, are integrated by parts in the component ``1'', {\it e.g.} 
$x_1 x_2 dx_1 =d(\frac{x_1^2 x_2}{2})-\frac{x_1^2}{2}dx_2$, leaving a 
``lower-order'' integral as above, plus an integral where the $dx_1$
differential has been traded for $dx_2$;
\item
an ordered double integral, whose integrand $h$ is symmetric, 
$h(x,y)=h(y,x)\,,$ amounts to one half of the corresponding unordered integral

\[
{\mbox{\bf P}}\!\int \!\int h(x,y) dx_2 dy_2=\frac{1}{2}\oint \oint 
h(x,y)dx_2 dy_2\,.
\]
\end{enumerate}

When all this has been done, we are left with a triple ordered integral 
with measure $dx_2 dy_2 dz_2$, whose integrand vanishes. A collection of double, 
non ordered integrals survives, but it can be shown it cancels as well.

The proof of invariance of $W[C]$ under $SL(2,R)$ is thereby completed.

\end{document}